 \documentclass[eps,12pt]{article}
 \usepackage{graphicx}
\begin{document}

\baselineskip=18pt

\begin{center}
{\Large\bf Solvation model dependency of helix-coil transition 
           in polyalanine}
\vskip 1.1cm
{\bf Yong Peng \footnote{E-mail:ypeng@mtu.edu} and
     Ulrich H.E. Hansmann \footnote{E-mail: hansmann@mtu.edu; to whom
     all correspondence should be addressed}}
\vskip 0.1cm
{\it Department of Physics, Michigan Technological University,
         Houghton, MI 49931-1291, USA}

\vskip 0.3cm
\today
\vskip 0.4cm
\end{center}
\begin{abstract}
Helix-coil transitions  in poly-alanine molecules of length 10 
are studied by  multicanonical Monte Carlo simulations. The solvation 
effects are included by either a distance-dependent dielectric 
permittivity or by a term that is proportional to the solvent-accessible
surface area of the peptide. We found a strong dependence of the
characteristics of the helix-coil transition from the details of
the solvation model.

\vskip 0.1cm
{\it Keywords:} Helix-coil transition, Protein folding, 
                Generalized ensemble simulations, solvation models

\end{abstract}



\section{Introduction} 
\vspace*{-0.5pt}
\indent
There has been recently a renewed interest in the conditions under which
\hbox{$\alpha$-helices}, a  common structure in proteins, 
are formed or dissolved.  It is long known that $\alpha$-helices 
undergo a sharp transition towards a random coil state when the 
temperature is increased. The characteristics of this so-called 
helix-coil transition have been studied extensively \cite{Poland}, 
most recently in Refs.~\cite{Jeff,HO98c}. In Refs.~\cite{AH99b,AH00b} 
evidence was  presented that the helix-coil transition in polyalanine
exhibits a true thermodynamic phase transition 
when interactions between all atoms in the molecule are taken into 
account \cite{AH99b,AH00b}. 

The later results were obtained from gas-phase 
simulations of poly-alanine. While there is some experimental 
evidence \cite{Jarrold} supporting the numerical results  of these
gas-phase simulations, the question remains how these results relate to
the biologically more relevant case  of solvated molecules. First 
investigations of this question were described in Refs.\cite{MO1,MO2} 
where it was claimed that the transition temperature is lower in water 
than in vacuum. However, that investigation relies on a single 
representation of the protein-water interaction and the dependence of 
their results on the details of the solvation term is not clear. 

In this paper,  we have  investigated how the 
characteristics of helix-coil transition change  with the details of 
the solvation term.  For this purpose, we have performed multicanonical 
simulations of polyalanine molecules of length 10. The protein-water 
interaction was included in two ways: either by  a distance-dependent 
dielectric permittivity or by a term that is proportional to the 
solvent-accessible surface area of the peptide.  For the later case 
we have considered four different parameter sets: OONS \cite{oons},
JRF \cite{jrf}, W92 \cite{wess} and  SCH \cite{sch}.  Quantities 
such as the energy, helicity and susceptibility were 
calculated as function of temperature. Our result were compared with 
that of gas phase simulations. A strong dependence of the characteristics 
of the helix-coil transition from the details of the solvation term was found.

\section{Methods} 
\vspace*{-0.5pt}
\noindent
Our investigation of the helix-coil transition for polyalanine is
based on a detailed, all-atom representation of that homopolymer. 
The interaction between the atoms  was
described by a standard force field, ECEPP/2,\cite{EC}  (as implemented 
in the program package SMMP \cite{SMMP}) and is given by:
\begin{eqnarray}
E_{tot} & = & E_{C} + E_{LJ} + E_{HB} + E_{tor},\\
E_{C}  & = & \sum_{(i,j)} \frac{332q_i q_j}{\epsilon r_{ij}},\\
E_{LJ} & = & \sum_{(i,j)} \left( \frac{A_{ij}}{r^{12}_{ij}}
                                - \frac{B_{ij}}{r^6_{ij}} \right),\\
E_{HB}  & = & \sum_{(i,j)} \left( \frac{C_{ij}}{r^{12}_{ij}}
                                - \frac{D_{ij}}{r^{10}_{ij}} \right),\\
E_{tor}& = & \sum_l U_l \left( 1 \pm \cos (n_l \chi_l ) \right).
\label{ECEPP/2}
\end{eqnarray}
Here, $r_{ij}$ (in \AA) is the distance between the atoms $i$ and $j$, and
$\chi_l$ is the $l$-th torsion angle. We have chosen ECEPP/2 instead
of the newer ECEPP/3 differs because this choice allows a more easy
comparison with our previous work. Both force fields differ from 
each other only in the way in which prolines and end groups are treated. 
In preliminary polyalanine simulations  we found no qualitative differences
in our results  when ECEPP/3 was used instead of ECEPP/2 (data not shown).

The interactions between our homo-oligomer and water are approximated by
means of two implicit water models. In the first model  (DDE)
the electrostatic interactions in the presence of water rely on 
a distance dependent electrostatic permittivity  \cite{hingerty}:
\begin{equation}
\label{eps}
\varepsilon(r) = D-\frac{D-2}{2}[(sr)^2+2sr+2]e^{-sr}~.
\end{equation}
For the parameters $D$ and $s$ empirical values are chosen such  that
for large distances the permittivity takes the value of  bulk water
($\varepsilon \approx 80$), and the value $\varepsilon=2$ for
short distances (protein interior space). Equation (\ref{eps}) is the 
result of interpolation of two types of interactions. For short 
distances it models the interaction of two charges placed in continuum 
medium, while over long distances it represents a Debye curve. This 
is clearly a gross  over-simplification  of protein-solvent interactions. 
However, approximating solvation effects by a distance-dependent dielectric
permittivity  was used by many authors to study the proteins and nucleic
acids (e.g. \cite{lavery}) since it does not significantly slow down 
protein simulations below that of simple {\it in vacuo} simulations.

In another common approximation of the protein-solvent interaction
one assumes that the
free energy contributions  from  atomic groups immersed in the
protein interior differ from  contributions of  groups exposed
to the water. It is commonly accepted \cite{oons,wess,lee,mclach}
that this free-energy difference is proportional to the
surface area of the atomic group which is  exposed to the solvent.
Within this approximation,  the total solvation energy of a protein is
given by the sum of  contributions from each solvated atomic groups:
\begin{equation}
\label{sigma}
E_{sol}=\sum_i\sigma_iA_i,
\label{Esol}
\end{equation}
where $E_{sol}$ is the solvation energy, $A_i$ is the conformational
dependent solvent accessible area of the surface of the $i-th$ atom
and $\sigma_i$ is the atomic solvation parameter for the atom $i$.
The summation is extended over all atomic groups.
The solvation parameters are  evaluated experimentally  by measuring
the free energy needed to bring the group from a nonpolar environment
(usually octanol or ethanol are used as convenient compounds) into water.
Many sets of solvation parameters were evaluated by several
authors with different methods, but unfortunately it is not always 
obvious which one is the most appropriate one. The sets we study here 
are named by us OONS \cite{oons}, 
JRF \cite{jrf}, W92 \cite{wess} and  SCH \cite{sch}, and are 
described in the respective references.

Simulations of detailed models of biological macromolecules are 
notoriously difficult. This is because the various competing interactions 
within the polymer lead to an energy landscape characterized by a 
multitude of local minima.  Hence, in the low-temperature region, 
canonical Monte Carlo or molecular dynamics 
simulations will tend to get trapped in one of these
minima and the simulation will not thermalize within the available
CPU time. Only recently, with the introduction of new and sophisticated 
algorithms such as  {\it multicanonical sampling} \cite{MU} and other 
{\it generalized-ensemble} techniques \cite{Review} was it
possible to alleviate this problem in  protein simulations \cite{HO}. 
For polyalanine,  both the failure of standard Monte Carlo techniques 
and the superior performance of the multicanonical algorithm are 
extensively documented  in earlier work \cite{OH95b}. For this reason,
we use again this sophisticated simulation technique for our project.  
 
In the multicanonical algorithm \cite{MU}
conformations with energy $E$ are assigned a weight
$  w_{mu} (E)\propto 1/n(E)$. Here, $n(E)$ is the density of states.
A  simulation with this weight
will  lead to a uniform distribution of energy:
\begin{equation}
  P_{mu}(E) \,  \propto \,  n(E)~w_{mu}(E) = {\rm const}~.
\label{eqmu}
\end{equation}
This is because the simulation generates a 1D random walk in the 
energy space,
allowing itself to escape from any  local minimum.
Since a large range of energies are sampled, one can
use the reweighting techniques \cite{FS} to  calculate thermodynamic
quantities over a wide range of temperatures $T$ by
\begin{equation}
<{\cal{A}}>_T ~=~ \frac{{\int dx~{\cal{A}}(x)~w^{-1}(E(x))~
                 e^{-\beta E(x)}}}
              {{\int dx~w^{-1}(E(x))~e^{-\beta E(x)}}}~,
\label{eqrw}
\end{equation}
where $x$ stands for configurations.

Unlike in the case of canonical simulations the weights 
\begin{equation}
w(E) = n^{-1}(E) = e^{-S(E)}
\label{eqweight}
\end{equation}
are not a priori known. Instead estimators for these weights have to
be determined. This is often done by an iterative procedure in which
for reasons of numerical stability Eq.~\ref{eqweight} is  replaced by
\begin{equation}
w(E) = e^{-\beta(E)E -\alpha(E)} ~.
\end{equation}
The multicanonical parameters $\beta(E)$ and $\alpha(E)$ are defined through
\begin{equation}
 \beta(E) = \frac{S(E') - S(E)}{E' - E}
\quad {\rm and} \quad
 \alpha (E) = \left\{ \begin{array}{ll}
                      0 &,~ E \ge E_{max}\cr
              \alpha (E') + (\beta (E') - \beta (E))E' &,~ E < E_{max}
                      \end{array} \right.
 \end{equation}
with $E$ and $E'$  adjacent bins in the array $ S (E)$. The 
$\beta(E)$ are then iteratively updated 
\cite{Berg} by the relation
\begin{equation}
\beta^{i+1}(E) = \beta^{i}(E) + g_0(E)\cdot \left(\ln H^{i}(E') 
- \ln H^{i}(E)\right)/(E'-E)~,
\label{berg1}
\end{equation}
in which $H^{i}(E)$ is the histogram of the $i$-th run (and $H(E) \ge 1$).
In Ref.~\cite{Berg} the  factor $g_0(E)$ in Eq.~\ref{berg1} was defined 
through
\begin{equation}
g_0(E) = \frac{\hat{g}^i(E)}{\hat{g}^i(E) + \sum_j^{i-1} \hat{g}^j (E)}
\quad {\rm with} \quad \hat{g}^i (E) 
= \frac{H^i(E')\cdot H^i(E)}{H^i(E')+H^i(E)}~.
\label{update}
\end{equation}
The above relation assumes that the histogram $H(E)$ counts independent
events which is in general not true. Hence, it is more appropriate and 
leads to a faster convergence of $\beta(E)$ if the array 
 $\hat{g}^i(E)$ in Eq.~\ref{update} is instead  defined by
\begin{equation}
\hat{g}^i(E) = \frac{K^i(E')K^i(E)}{K^i(E')+K^i(E)}
\end{equation}
where the auxiliary array $K(E)$ now counts  only the number 
of {\it independent} visits at energy $E$.

With the above described iterative procedure we needed  200,000 sweeps for
the weight factor calculations. All thermodynamic quantities were  then
estimated from one production run of $1,000,000$ Monte Carlo sweeps 
starting from a random initial conformation, i.e. without introducing 
any bias.

\section{Results and Discussion}
In previous gas-phase simulations of poly-alanine 
\cite{HO98c,AH99b,AH00b,OH95b}
we observed at $T=430\, K$ a pronounced transition between a high-temperature 
phase dominated by disordered coil structures and an ordered phase
with single, extended helices. A natural order parameter 
for this helix-coil transition is the average number $<n_H(T)> $
of  residues in the oligomer which are part of an
$\alpha-$helix. Following earlier work \cite{OH95b} we define a residue
as helical if the pair of backbone dihedral angles $\phi,\psi$ takes a
value in the range $(-70\pm 20,-37\pm 20)$. In Fig.~1a  this order 
parameter is displayed as function of temperature for a gas-phase
simulation (GP) of Ala$_{10}$ and  simulations with the various solvation
terms. Fig.~1b shows the corresponding plots for the susceptibility
$\chi (T)$ defined by
\begin{equation}
\chi(T) = <n_H^2 (T)> - <n_H(T)>^2 \quad .
\end{equation}

In Fig.~1a and 1b the curves, representing the various simulations,
fall into three groups. For the case where the protein-solvent
interaction was approximated by a distance-dependent permittivity 
(DDE), both $<n_H>$ and $\chi$  have a similar temperature dependence
than is observed for poly-alanine in gas-phase simulations (GP).
However, the transition temperature $T_c$ is shifted from
$T=435\pm 20$ K (gas-phase) to a {\it higher} value $T=495\pm 20$.
This temperature was determined from the maximum of the 
susceptibility $\chi (T)$ in Fig.~1b
and  is listed in table~1. To the same group belong the simulations
in which the solvation energy was approximated by a solvent 
accessible surface term  with either the OONS \cite{oons} or 
SCH \cite{sch} parameter set. In both cases susceptibility $\chi$
and  order parameter  $<n_H(T)>$ show also a  temperature dependence
similar to  the one of  gas-phase simulations. Only now, the transition
temperature $T_c$ is shifted to {\it lower} temperatures. The corresponding
transition temperatures can be again determined from the positions
of the maximum in $\chi(T)$ and are also listed in table~1.
The  shift towards lower temperatures was one of the main results
reported in Refs.~\cite{MO1,MO2} for simulations with the OONS 
solvation energy, and our $T_c=345\pm 20$ K agrees well with their value
$T_c= 340$ K (no errors quoted) in Refs.~\cite{MO1,MO2}.

A somehow different behavior is observed in the simulation where
the protein-water interaction was approximated by a solvent
accessible surface term relying on the W92 \cite{wess} parameter set.
Here, the form of $<n_H>$ indicates only partial helix formation 
and  occurs only  at much lower temperatures.
The susceptibility  $\chi(T)$ in Fig.~1b gives no indication for
a helix-coil transition. For this reason no value of $T_c$ is listed
for the W92 parameter set in table~1. Instead, we observe in Fig.~2 
for this case at low temperatures  
even the appearance of residues whose backbone 
dihedral angles $\phi,\psi$ take values typical for a $\beta$-sheet  
$(-150\pm 30,150\pm 30)$.  

Yet another behavior is observed in simulations where 
the solvation energy of Eq.~\ref{Esol} is evaluated by means  
of the JRF parameter set. No formation of helices or sheets 
is observed in Figs.~1 and 2. Since  no transition temperature
can be determined, we do not list a value of $T_c$ for
the JRF parameter set in table~1. 

The same grouping can be found in Fig.~3a-f  where we display 
various energy terms as a function of temperature.  In these 
figures we have shifted the  solvation energies and the 
partial ECEPP/2 energies $E_{C}, E_{LJ}, E_{HB}$ 
and $E_{tor}$ of Eq.~\ref{ECEPP/2}
by a constant term  such that we have for all  solvation models  at $T=1000$ K 
 $E_{sol} = 0$ and $E_C=E_{LJ}=E_{HB}=E_{tor}=0$. Such a shift 
by an irrelevant constant allows  a better comparison of the different
simulations. The average total energy $<E_{tot} > $ 
which is the sum of intramolecular potential energy $E_{ECEEP/2}$
and  the solvation energy $E_{sol}$, is displayed in Fig.~3a.
We observe  again that simulations with the parameters sets OONS and SCH,
and such with distant dependent permittivity (DDE), have a similar 
temperature dependence as gas phase simulations (GP). On the other hand,
in simulations relying on the W92 parameter set, the energy varies 
less with temperature and is  at low temperatures considerably higher
than  in the simulations with  other  solvation energy terms. Finally,
the energy in simulations with the JRF parameter set is an almost 
linear function of temperature and is especially at high temperatures 
much lower than the energies found in  gas phase simulations.

The dissimilar behavior of energy for simulations with different
solvation terms is even more obvious in Fig.~3b where the average 
intramolecular energy $E_{ECEPP/2}$ is drawn. While this energy term 
decreases between $1000$ K and $150$ K by $\approx 50$ Kcal/mol 
(with most of that change, $\approx 30$ Kcal/mol, happing around 
the respective transition temperature $T_c$) in
gas-phase simulations (GP) and in simulations with OONS, SCH and DDE 
solvation terms, it changes in the same temperature interval only
by $\approx 20$ Kcal/mol   in simulations utilizing the JRF or W92
parameter sets. Since for these two parameter sets  also 
no or only little helix formation was observed  it seems likely that 
the formation of helices 
is related to the large gain in potential energy observed for GP,
OONS,SCH and DDE simulations. This gain in potential energy is in part due
to the formation of hydrogen bonds  between a residue and the fourth 
following one in the polypeptide chain which stabilize an $\alpha$-helix. 
Fig.~3c displays
the average hydrogen-bonding energy $<E_{HB}>$ of Eq.~\ref{ECEPP/2} 
as a function of temperature and one can clearly see the gain in energy for
the GP, DDE, OONS and SCH simulations at the respective
helix-coil transition temperatures of table~1. No such gain is observed
in W92 and JRF simulations where also no helix formation was found. A
similar gain in energy with helix formations in gas-phase and simulations
with DDE,OONS and SCH solvent representations is also observed for
the average Lennard-Jones energy $<E_{LJ}>$ and the electrostatic 
energy $<E_C>$ displayed in Fig.~3d and 3e, respectively. Note also in
Fig.~3e the large gain in $E_C$ for DDE at the helix-coil transition 
temperature which additionally stabilizes the $\alpha$-helix in this model.

A complementary picture is found in Fig.~3f where the solvation
energy $E_{sol}$ is shown as a function of temperature. The 
observed helix formation in gas phase simulations and such with 
OONS,SCH and DDE solvent representations is correlated with 
an increases of the  solvation energies by $\approx 5$ Kcal/mol.
On the other hand, in simulations with the W92 and JRF parameter sets, 
for which no helix-formation was observed in Fig.~1,  $E_{sol}$ decreases 
with temperature. 
This decrease is only  $\approx 5$ kcal/mol for W92, but it is much larger
(of order $30$ kcal/mol) in the case of JRF where the solvation energy 
is the dominant term.

The effects of the dominant solvation term in simulations with the JRF
parameter set can also be seen in Fig.~4. In this figure  the average 
radius of gyration, a measure for the compactness of configurations,
is shown as a function temperature. One can see that this quantity 
changes little with temperature for the JRF data. However, its value is
over the whole temperature range considerably smaller than observed
in the other simulations. 
This indicates that the JRF solvation term  favors already at
high temperatures  compact configurations,
 and that the pressure towards compact structure is such
that the more elongated helices cannot be formed. Note however, that
the tendency towards compact configurations does not lead to a lower
Lennard-Jones energy $E_{LJ}$ as one would expect. Fig.~3d indicates
that $<E_{LJ}>$ is at low temperatures in JRF simulations even larger
than in GP, DDE, OONS and SCG simulations where helix-formation was observed.
The tendency towards compact structures in JRF simulations may be
due to the fact that JRF parameter set was developed from minimum energy
(i.e. compact) conformations of peptides (the low-energy conformations of 13 
tetrapeptides derived by NMR studies \cite{jrf}), and therefore 
this parameter set may have 
an intrinsic bias towards compact structures. 

On the other hand, the W92
parameter set was developed from measurements of 
free energies of amino acid side
chain analogs from vapor to water \cite{Wolfenden}. The parameters for this
set are negative for all atoms except carbon meaning that the nitrogen,
oxygen and sulfur atoms are considered hydrophilic, i.e. favoring
solvent exposure. This explains not only the small solvation energies
observed for this parameter set in Fig.~3f, but also why in Fig.~4
the radius of gyration is consistently larger for this parameter set
than for the others indicating that extended configurations are favored 
with this parameter set. This bias towards extended structures 
limits  again  the formation of $\alpha$-helices. 

While the OONS parameter set was derived from experimental
free energies of gas-to-water transfer of small aliphatic and aromatic 
molecules, the SCH is not directly based on experimental free energy data.
Instead, it was developed as an optimized parameter set to complement the
CHARMM force field \cite{charmm}. In both  parameter sets  the hydrophobic
character of the carbon atoms is increased and the hydrophilic character
of uncharge oxygen and nitrogen atoms decreased resulting into the large
solvation energies of these two parameter sets (when compared with the 
one of the W92 parameter set) that one observes in Fig.~3f. The OONS and 
SCH solvation energies again favor extended structures (the radius of 
gyration has larger values than found in gas-phase simulations), however, 
the interplay of solvation energies and
intramolecular ECEPP/2 energy is such that the radius of gyrations
(and consequently the compactness) of polyalanine configurations as
a function of temperature shows a similar behavior as the gas-phase
simulation. The same is true for the DDE simulation where the protein-solvent
interaction was approximated by a distance-dependent permittivity.

Our results demonstrate that the helix formation is due to the gain in
potential (intramolecular) energy while (with the exception of the 
JRF parameter set)  the solvent-accessible surface terms favor 
extended peptide configurations. Table 2 summarizes the differences in total
energy $\Delta E_{tot}$ , solvation energy $\Delta E_{sol}$, potential
energy $\Delta E_{ECEEP/2}$, and the partial energies $\Delta E_C$,
 $\Delta E_{LJ}$, $\Delta E_{HB}$ and $\Delta E_{tor}$  between
complete helical configurations (all residues with exception of
the terminal ones are part of an $\alpha$-helix)
and coil configurations at temperature $T = 300$ K for  gas-phase, DDE
OONS and SCH simulations. Note, that the intramolecular
energy differences $\Delta E_{ECEPP/2}$ of gas-phase, 
 OONS and SCH simulations have within their error bars the same values.
 For simulations with the W92 parameter set 
the longest found helix consists of 6 consecutive residues. Hence, we
measured for this case only the energy difference between configurations
with at least three consecutive helical residues (i.e. one turn of
an $\alpha$-helix) and coil configurations. This modified definition
of the energy differences is also the reason for
the smaller value of $\Delta E_{ECEPP/2}$ listed for W92 in table~2.
We do not list  energy differences for the JRF parameter set since 
no helices were found in simulations utilizing this parameter set. 

Note that in  simulations with distant dependent permittivity (DDE) 
helices are energetically more favored than in the gas-phase simulations. 
This is due to the increased contribution from the Coulomb term $E_C$ 
as one can also see in Fig.~3e. The larger energy gap between
helical and coil conformations (when compared with gas-phase 
simulations)   explains  why the transition temperature is higher
in DDE simulations than in gas-phase simulations.

For the OONS and the SCH
parameter set the solvation energy difference $\Delta E_{sol}$ 
is positive (indicating that coil structures are energetically favored), 
but its magnitude is only approximately half that of the potential 
energy difference $\Delta E_{ECEPP/2}$. Hence, there is still an overall 
energetic gain connected with helix formation. However, in both cases
the total energy difference between helical and
coil configurations is reduced by the solvation energy  when compared
with the gas-phase simulation. This reduction of the energy gap leads
to the lower transition temperatures observed in OONS and SCH simulations.

On the other hand, for the W92 parameter set  
we find that $\Delta E_{ECCP/2}$ and $\Delta E_{sol}$
are of same magnitude so that helical configurations are not or only weakly
energetically favored. This is consistent with our results in Fig.~1a and 1b
where we  find at $T=280$ K a high average helicity in OONS and SCH 
simulations but only a small value of $<n_H>$ and no indications
for a helix-coil transition in W92 simulations.  An evaluation
of energy differences was not possible for simulations with
the JRF parameter set since no helices were found.

The above results indicate that the existence and characteristics of
the helix-coil transition in polyalanine depend strong on the details of
the solvent representation. In order to evaluate the validity of the
different solvent models one has to compare the numerical results with
experimental data. For this purpose we have calculated the helix
propagation parameter $s$ which was also determined by experiments 
\cite{WAS,CB}.  According to the Zimm-Bragg model \cite{ZB}  the 
average number of helical residues $<n>$ and the average length $<\ell>$ 
of a helical segment are given for large number of residues $N$  by 
\begin{eqnarray}
{{<n>} \over N}~ &=& ~{1 \over 2} - {{1-s} \over {2
\sqrt{(1-s)^2 + 4s \sigma}}}~, \\
<\ell>~~ &=& ~1 + {2s \over {1-s+\sqrt{(1-s)^2 +4s \sigma}}}~,
\end{eqnarray}
where $s$ is  the helix propagation parameter and  $\sigma$ the 
nucleation parameter of the Zimm-Bragg model.
From these equations with the values of ${<n>}/ N$ and $<\ell>$
calculated from the multicanonical production runs, we have calculated
$s$ at temperature $T=280$ K for gas-phase and the different solvation
models. Our values are summarized in table 3 which also lists  our
 $\sigma$ values.  Our results for gas-phase, DDE and OONS simulations 
are in agreement with the experimental results of Ref.~\cite{CB}
 where they list values of $s$(Ala) between $1.5$ and $2.19$. 
On the other hand, the 
$s$ value obtained in the SCH simulation  agrees well with the one
obtained by the host-guest technique of Ref.~\cite{WAS}. However,
the $s$ values which were obtained in W92 or JRF simulations do not
agree with either of the experimental data. Hence, we conclude
that the W92 and JRF parameter sets are not appropriate solvation
models in simulations of polyalanine. Otherwise, the variation
in the experimental data is too large to give indications whether
 one of the remaining solvent representations  (DDE, OONS, SCH or
even no solvent at all (GP)) is preferable over  the others.

\section{Conclusions}
We have performed multicanonical simulations of polyalanine. The
intramolecular forces were modeled by the ECEPP/2 force field and
various approximations for the solvation energy were studied. 
We observed that whether a helix-coil transition is observed 
for poly-alanine, and at what temperature, depends strongly 
on the chosen approximation for the protein-solvent interaction. 
Our results demonstrate both the importance (and need) of including 
solvation terms into protein simulations and the difficulties in 
chosing an adequate representation of the protein-water interactions.
Especially when using the solvent-accessible surface approach, it seems
necessary to carefully choose a parameter set  
that is suitable for the problem under consideration. Use of a specific
parameter set without further justification  could otherwise
generate  miss-leading  results.

\section*{Acknowledgement}
 U. Hansmann gratefully acknowledges support by  a research grant 
from the National Science Foundation (CHE-9981874). This article was
written in part while U.H. was visitor at the Department of Physics at
University of Central Florida. He thanks  Alfons Schulte, Weili Luo, 
Aniket Bhattacharya and Brian Tonner for their kind hospitality
during his stay in Orlando.



\newpage
{\huge Tables:}
\begin{table}[h]
\caption{Transition temperatures for the helix-coil transition
         in ALA$_{10}$ as obtained from gas-phase simulations 
         and simulations with various solvent representations.
         All results rely on multicanonical simulations of 1,000,000
         Monte Carlo sweeps each.}
\begin{tabular}{cc}
\hline
   Model          & $T_c$        \\
   GP &  435(20)  \\
   DDE            &  495(20) \\
   OONS           &  345(15) \\
   SCH            &  285(25) \\
   W92            &   - \\
   JRF            &   - \\
\hline
\end{tabular}
\end{table}
\begin{table}[h]
\caption{Energy differences between helical and
         configurations (see text) at $T=280$ K as measured in
         gas-phase simulations and simulations with various
         solvent representations. All results rely on multicanonical 
         simulations of 1,000,000 Monte Carlo sweeps of ALA$_{10}$
         for each case.}
\begin{tabular}{cccccccc}
\hline
 Model & $\Delta E_{tot}$ & $\Delta E_{sol}$ & $\Delta E_{ECEPP/2}$ &
        $\Delta E_{C}$& $\Delta E_{LJ}$ & $\Delta E_{HB}$ & $\Delta E_{tor}$\\
\hline
 GP  &$-16.9(1)$ & -  &$-16.9(1)$ &$0.4(3)$ &$-12.1(1)$ &$-4.3(3)$ &$-0.8(1)$\\
DDE&$-17.9(6)$& -  &$-17.9(6)$ &$-3.6(2)$&$-10.1(4)$&$-3.9(2)$&$-0.3(1)$\\
OONS&$-11.3(9)$&$4.1(3)$&$-15.4(6)$&$-0.2(1)$&$-10.7(4)$&$-4.1(1)$&$-0.4(1)$\\
SCH&$-7.1(5)$&$8.7(1)$&$-15.8(5)$&$0.7(3)$&$-11.2(2)$&$-4.6(3)$&$-0.7(1)$\\
W92&$-0.7(7)$&$5.6(6)$&$-6.3(1.1)$&$0.8(1)$&$-5.8(9)$&$-1.0(2)$&$-0.3(1)$\\
JRF& --      &   --   &   --      & --     &   --    &  --     &  --    \\
\hline
\end{tabular}
\end{table}

\newpage
\begin{table}[h]
\caption{Helix propagation parameter $s$ and nucleation parameter $\sigma$
         at $T=280$ K for Ala$_{10}$ as measured in
         gas-phase simulations and simulations with various
         solvent representations. All results rely on multicanonical
         simulations of 1,000,000 Monte Carlo sweeps for each case.}
\begin{tabular}{ccc}
\hline
 Model &  $s$ &  $\sigma$ \\
\hline
 GP    & 1.67(9) & 0.15(1)\\
 DDE   & 1.78(12)& 0.15(1)\\
 OONS  & 1.31(15)& 0.13(1)\\
 SCH   & 1.02(15)& 0.11(2)\\
 W92   & $\approx 0$ & $ > 1$ \\
 JRF   & $\approx 0$ & $ > 1$ \\
\hline
\end{tabular}
\end{table}
\vfill
\newpage

\pagebreak
{\huge Figure Captions:} \\
\begin{description}
\item[Fig.~1] Temperature dependence of (a) the average number 
               $<n_H>$ of helical residues and (b) the
               susceptibility $\chi (T)$ for ALA$_{10}$ 
               as calculated from a gas-phase
               simulation and from simulations with various 
               solvation energy terms. All results rely on
               multicanonical simulations of 1,000,000 Monte 
               Carlo sweeps each.
\item[Fig.~2]  Temperature dependence of the average number 
               $<n_B>$ of residues whose backbone dihedral angles
               $\phi,\psi$ take values as typically found in
               $\beta$-sheets. Results from a gas-phase simulation
               and such with various solvation terms are presented
               for ALA$_{10}$.
               All data rely on multicanonical simulations of
               1,000,000 Monte Carlo sweeps.
\item[Fig.~3]  Temperature dependence of (a) the total energy
               $<E_{total} = E_{ECEEP/2} + E_{sol}>$, (b) the
               intramolecular energy $<E_{ECEPP/2}>$,  (c) the
               hydrogen-bonding energy $<E_{HB}>$, (d) Lennard-Jones
               energy $<E_{LJ}>$, (e) Coulomb energy $<E_C>$,  and (f) the
               solvation energy $<E_{sol}>$ as calculated from a gas-phase
               simulation and from simulations with various
               solvation energy terms. All results rely on
               multicanonical simulations of ALA$_{10}$ with 1,000,000 Monte
               Carlo sweeps for each case.
\item[Fig.~4]  Temperature dependence of the average radius-of-gyration
               $<R_{gy}>$ as measured in gas-phase simulations and
               simulations with various solvent representations.
               All data rely on multicanonical simulations of
               1,000,000 Monte Carlo sweeps.
\end{description}

\vfill



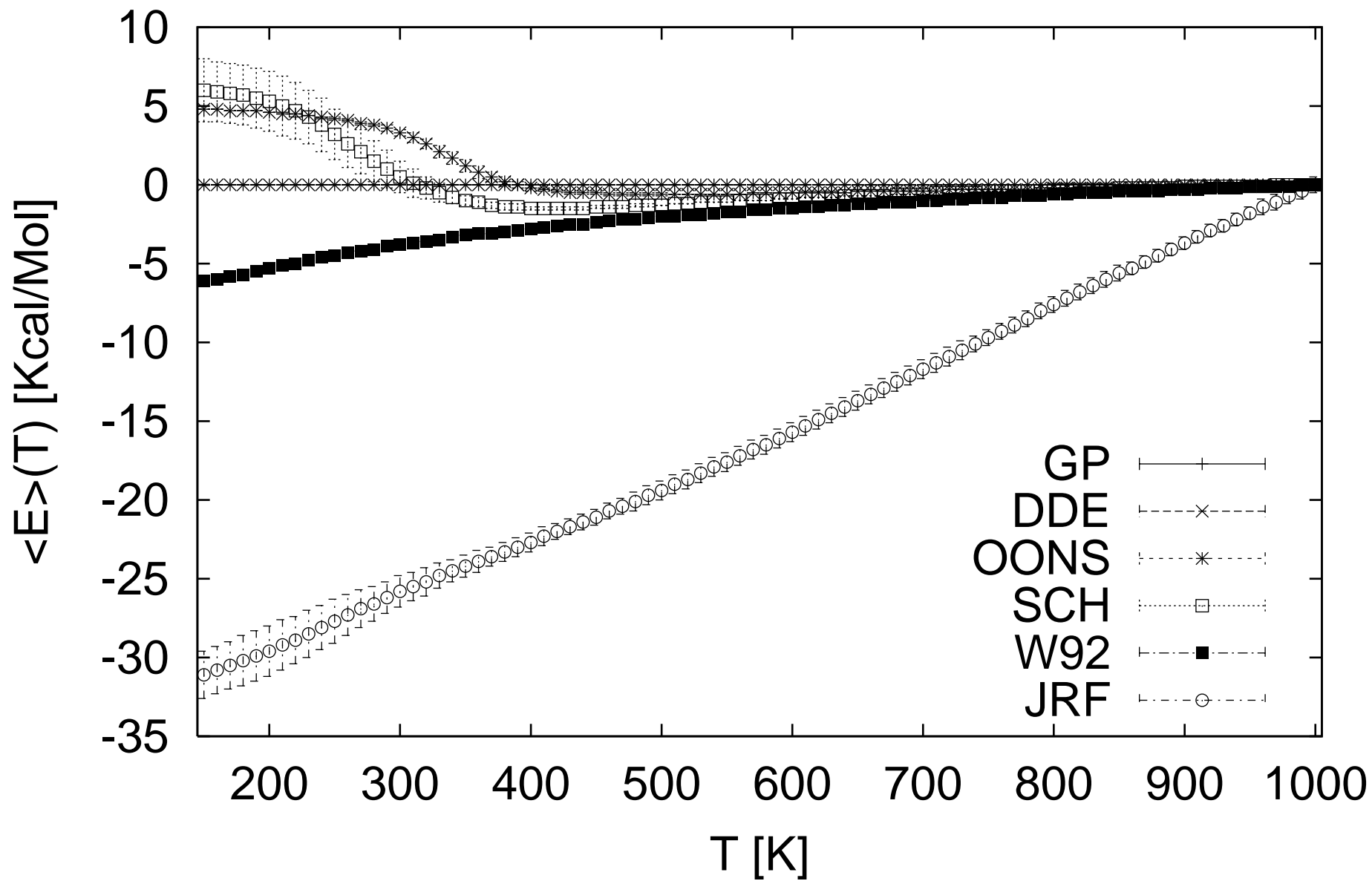
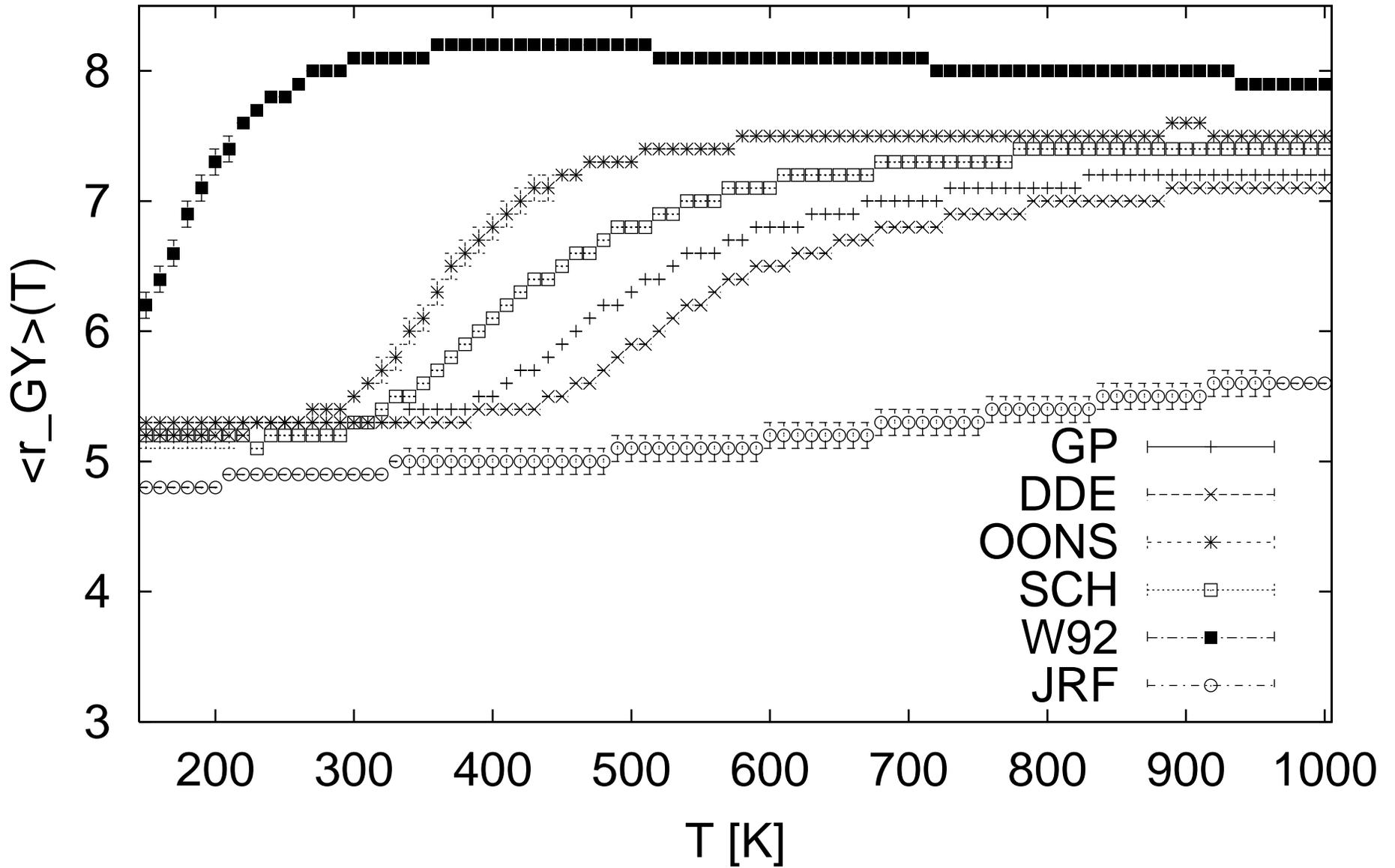
\end{document}